\documentstyle[twocolumn,prl,aps]{revtex}
\newcommand{\be}{\begin{equation}} \newcommand{\ee}{\end{equation}} 
\newcommand{\bea}{\begin{eqnarray}}\newcommand{\eea}{\end{eqnarray}}
\newcommand{\bm}[1]{\mbox{\boldmath$#1$}}
\newcommand{\grad}{\bm \nabla}
\begin{document}
\draft
\preprint{OCHA-SP-01-09, cond-mat/0111523}
\title{ Exact results on the dynamics of multi-component Bose-Einstein
condensate}
\author{Pijush K. Ghosh}
\address{ 
Department of Physics,
Ochanomizu University,\\
2-1-1 Ohtsuka, Bunkyo-ku,
Tokyo 112-8610, Japan.\\}

\maketitle
\begin{abstract} 
We study the time-evolution of the two dimensional multi-component
Bose-Einstein condensate in an external harmonic trap with arbitrary
time-dependent frequency. We show analytically that the time-evolution
of the {\it total mean-square radius} of the wave-packet is determined in
terms of the same solvable equation as in the case of a single-component
condensate. The dynamics of the {\it total mean-square radius} is also the
same for the rotating as well as the non-rotating multi-component condensate.
We determine the criteria for the collapse of the condensate at a finite time.
Generalizing our previous work on a single-component condensate, we show
explosion-implosion duality in the multi-component condensate.
\end{abstract}
\narrowtext

The successful creation and observation of Bose-Einstein condensation(BEC)
in dilute alkali atoms have opened up a plethora of new possibilities to test,
otherwise intractable, many-body quantum phenomenon in the
laboratory\cite{rv1}. The Gross-Pitaevskii equation(GPE), the mean-field
description of the BEC, is successful enough in explaining
most of the observed results as well as predicting new phenomenon. The methods
involved in studying the GPE are mainly numerical and/or approximate:
perturbative
and variational. The exact and analytical results of a nonlinear equation, if
known, not only act as a guide to determine the validity of different
approximate and numerical methods; they also give rise to new, counter-intuitive
results in some cases. Unfortunately, no exact solution of GPE is known except
for in one dimension.

The two dimensional GPE, like its counterparts in higher dimensions, is not
exactly solvable. However, due to an underlying dynamical $O(2,1)$
symmetry\cite{O21},
the time-evolution of certain moments related to the two dimensional GPE
can be described exactly\cite{ami}. This result is valid even if the
condensate is considered in a time-dependent harmonic trap. This leads to the
prediction of explosion-implosion duality\cite{mai} and extended parametric
resonance\cite{mai,spain} in the two dimensional BEC. Both of these phenomenon
are universal for any non-relativistic theory having dynamical $O(2,1)$
symmetry\cite{ami,mai,tkg}. Interestingly enough, apart from the two
dimensional BEC, the same explosion-implosion duality can also be observed in
supernova explosion and in laser induced implosion in
plasma\cite{astro,astro1}. This shows the importance of exact methods,
based on an underlying symmetry, in relating diverse areas of physics such as
the BEC and the supernova explosion.

The results described above are for a single-component condensate, where the
spin degree's of freedom have been frozen though the use of a magnetic trap.
Recently, the spinor condensate with independent spin degree's of freedom has
also been created and observed in the laboratory\cite{exp}. Similarly,the
two-component condensate, where two different hyperfine states of the same
atomic species are condensed simultaneously, has also been experimentally
realized\cite{2comp}. The spinor condensate\cite{oka,ho,rv2,rv3} and the
two-component condensate\cite{shen} have a very rich structure compared to
the single-component condensate. This is manifested in the existence of
topological defects like skyrmion, domain-wall, vortices and Alice string in
these condensates\cite{taku}.

The purpose of this note is to extend the studies of Refs. \cite{ami,mai}
on the two
dimensional single-component BEC to the two dimensional multi-component
BEC. The experimentally realizable two-component and the spinor condensate can
be obtained as special cases of this general multi-component BEC. We study the
exact time-evolution of the second moment of the two dimensional
multi-component condensate in an arbitrary time-dependent harmonic trap.
This particular second moment can be identified as the {\it total mean-square
radius} of the condensate. We show that the dynamics of the second moment is
determined by the same solvable equation, as in the case of a single-component
condensate. No matter how many components are there, or how they interact among
themselves, or even whether they are rotating or non-rotating, the dynamics of
the {\it total mean-square radius} is universally determined by the same
equation. The detail information on the system is encoded, through the
Hamiltonian, into a constant of motion appearing in this universal equation.
Thus, the dynamics of the system can be studied in terms of the same set
of initial conditions for any number of components. We determine the criteria
for the collapse of the condensate of this system. We also show that the
multi-component BEC, in its full generality, exhibits an explosion-implosion
duality and extended parametric resonance for special choices of the
time-dependence of the trap. All these results are exact and analytical.

Consider the following Lagrangian in $2+1$ dimensions,
\bea
{\cal{L}} & = & \sum_a \left ( i \psi_a^* \partial_{\tau} \psi_a
- \frac{1}{2 m} {\mid {\bf \grad} \psi_a \mid}^2 \right )\nonumber \\
&& - \frac{1}{2} \sum_{abcd} g_{abcd} \ \psi_a^* \psi_b^* \psi_c \psi_d,
\ \ a, b, c, d = 1, 2, \dots, n,
\label{eq1-}
\eea
\noindent where $n$ is the total number of components. The coupling constants
$g_{abcd}$ are related to the s-wave scattering length matrix. The possible
values of $g_{abcd}$, and hence of the scattering length matrix, may be
constrained by symmetry requirements. For example, the special case of
a two-component condensate can be obtained by choosing $n=2$ and
$g_{abcd}= \frac{1}{2} ( \delta_{ac} \delta_{bd} + \delta_{ad} \delta_{bc}
) \bar{g}_{ab}$ so that the system has a global $U(1)^2$ symmetry. 
A phase-separation occurs for such a system if all the scattering lengths
are positive and satisfy the inequality, $g_{12}^2=g_{21}^2 > g_{11} g_{22}$
\cite{shen}.
Similarly, the spin one spinor condensate can be obtained by choosing $n=3$ and
$g_{abcd}= \frac{1}{2} \left [ g_1 \delta_{ac} \delta_{bd} +
g_2 \sum_{\alpha} ( S_{\alpha} )_{ac} (S_{\alpha})_{bd} + ( a \leftrightarrow b)
\right ]$, where $S_{\alpha}$'s are three spin-matrices. A positive
$g_2$ defines an anti-ferromagnetic regime, while the ferromagnetic regime
is characterized by a negative $g_2$. It is known that the ferromagnetic
or the anti-ferromagnetic nature of the interaction plays an important role
to characterize different properties of the condensate\cite{oka,ho}.
Both the phenomenon of the phase separation and the ferromagnetic or
anti-ferromagnetic nature of the ground state are specific to multi-component
condensate for $n \geq 2$. Further, note that we have additional terms
describing the interaction among different components
as we go from the single-component to the two-component, to spinor and to the
general multi-component condensate described by (\ref{eq1-}).
However, to our surprise, the dynamics of the {\it total mean-square radius}
is independent of such variation in the inter-component interaction and
universally determined by the same solvable equation as in the case of a 
single-component($n=1$) condensate. Consequently, the criteria for the collapse
of the condensate at a finite time is also the same for any $n$-component
condensate. For the very special case of an additional global $U(1)^n$ 
symmetry in (\ref{eq1-}), such a result has been obtained previously in Ref.
\cite{tsu}. We remark that our results are much more general. Moreover, the
known results are reproduced in a very elegant way. We will consider only the
most general form of ${\cal{L}}$ from now onwards, since our result is
independent of particular details of the interaction.

All the coupling constants $g_{abcd}$ have the inverse-mass dimension in the
natural units with $c=\bar{h}=1$. This allows to have a scale and conformally
invariant theory. The action 
${\cal{A}} =\int d \tau d^2{\bf r} {\cal{L}}$ is invariant under the following
time-dependent transformations\cite{trans,rj,ch,si,pla,st,me},
\bea
&& {\bf r} \rightarrow {\bf r_h} = {\dot{\tau}}(t)^{-\frac{1}{2}} {\bf r}, \ \
\tau \rightarrow t = t(\tau), \ \ \dot{\tau}(t) =
\frac{d \tau(t)}{d t}, \nonumber \\
&& \psi_a(\tau, {\bf r}) \rightarrow \psi_a^h(t, {\bf r}_h) =
\dot{\tau}^{\frac{d}{4}} exp \left ( - i m \frac{\ddot{\tau}}{4 \dot{\tau}}
r_h^2 \right ) \psi_a(\tau, {\bf r}),
\label{eq5}
\eea
\noindent with the scale-factor $\tau$ given by,
\be
\tau(t) = \frac{ \alpha t + \beta}{\gamma t + \delta}, \ \
\alpha \delta - \beta \gamma =1.
\label{eqss}
\ee
\noindent Note that all the components of the order parameter are
multiplied by the same time-dependent scale-factor and the phase in the
symmetry-transformation above. One might naively think that the requirement
of the identical phase-factors for all the components of the order parameter
is due to the interaction term. However, this is not the case.
Even if we consider the free theory, i.e. $g_{abcd}=0$ for all values of
the indices, the requirement of the identical phases in (\ref{eq5}) is
essential in order it to be a symmetry transformation. This is precisely
because the transformation of the scalar-fields is coupled with that of
the space-time coordinates. If we choose some special values for the coupling
constants $g_{abcd}$ such that the Lagrangian has an internal global symmetry,
say for example $SU(n)$, we certainly have the freedom of varying the 
phase-factors up to a global $SU(n)$ rotation. However, such an additional
internal symmetry are completely decoupled from the symmetry
transformations described in Eqs. (\ref{eq5}) and (\ref{eqss}), and do not
have any effect on our results.

Let us now introduce two moments $I_1$ and $I_2$ in terms
of the density $\rho$ and the current ${\bf j}$ as,
\bea
&& \rho (\tau, {\bf r}) = \sum_a \psi_a^* \psi_a,\nonumber \\
&& {\bf j}(\tau, {\bf r}) = -\frac{i}{2 m} \sum_a \left ( \psi_a^* 
{\bf \grad} \psi_a - \psi_a { \grad}
\psi_a^* \right ),\nonumber \\
&& I_1 (\tau) = \frac{m}{2} \int d^2 {\bf r} \ r^2 \ \rho,\nonumber \\
&& I_2 (\tau) = \frac{m}{2} \int d^2 {\bf r} \ {\bf r} \cdot {\bf j}.
\label{eq2}
\eea
\noindent We are dealing with a conservative system and the total number
of particles $ N(\tau)= \int d^2 {\bf r} \rho$ is a constant of motion.
The global $U(1)$ symmetry of ${\cal{L}}$ can be enlarged to $U(1)^n$ for
certain special choices of $g_{abcd}$. The total number of particles for
each species are conserved separately for this case. However, as emphasized
earlier, such an additional internal symmetry do not have any significant
effect on our results. Thus, only the conservation of the total number of
particles $N$ is important for our study.
The moment $I_1$ is the sum of the mean-square radii corresponding to each
and every components. This moment can be interpreted as the square of the
width of the wave-packet for the single-component condensate, when confined
in an external harmonic trap\cite{spain}. However, for the multi-component
case, the moment $I_1$ can not be identified as the total width of the
wave-packet. As emphasized in our previous work\cite{ami}, the moment $I_1$
has been used extensively in the analysis of the non-linear
Schr$\ddot{o}$dinger equation( NLSE) \cite{spain,tkg,za,mw,bull},
BEC\cite{gp} and in optics\cite{opt}. The dynamics of $I_1$,
when the system (\ref{eq1-}) is immersed in an external time-dependent
harmonic trap, is the central subject of the investigation of this letter.
We show that the dynamics of $I_1$ is universally determined by the same
solvable equation, as in the case of a single-component BEC.

Particular choices of $\tau(t)=  t + \beta, \alpha^2 t$,
and $ \frac{t}{1 + \gamma t}$, correspond to time translation, dilation
and special conformal transformation. The corresponding generators of these
transformations, the Hamiltonian $H$, the dilatation generator $D$ and the
conformal generator $K$ are,
\bea
&& H = \int d^2 {\bf r} \left [ \frac{1}{2 m}
\sum_a {\mid \grad \psi_a \mid}^2 +  
\frac{1}{2} \sum_{abcd} g_{abcd} \ \psi_a^* \psi_b^* \psi_c \psi_d
\right ],\nonumber \\
&& D =  \tau H - I_2,\nonumber \\
&& K = - \tau^2 H + 2 \tau D + I_1.
\label{eq3}
\eea
\noindent 
These generators close under the algebra,
\be
[H, D] = i H, \ \ [H, K] = 2 i D, \ \ [K, D] = -  i K,
\label{eq3.2}
\ee
\noindent if we promote the fields $\psi_a$'s to the operators $\hat{\psi}_a$
with the following bosonic commutation relations among themselves,
\bea
&& \left [ \hat{\psi}_a ({\bf r}), \hat{\psi}_b^* ({\bf r}^{\prime}) \right ] = 
\delta_{ab} \ \delta(\bf{r} - {\bf r}^{\prime}),\nonumber \\
&& \left [ \hat{\psi}_a ({\bf r}), \hat{\psi}_b ({\bf r}^{\prime}) \right ] =
\left [ \hat{\psi}_a^* ({\bf r}), \hat{\psi}_b^* ({\bf r}^{\prime}) \right ]
= 0.
\label{eq3.1}
\eea
\noindent The algebra given by Eq. (\ref{eq3.2}) defines a conformal group,
which is isomorphic to the group $O(2,1)$\cite{trans}. Thus, the system
(\ref{eq1-}) has a dynamical $O(2,1)$ symmetry with the interpretation of the
fields $\psi_a$'s as the operators $\hat{\psi}_a$ satisfying (\ref{eq3.1}).
In this note, we will be considering only the fields
$\psi_a$'s, not the operators $\hat{\psi}_a$. We do not make use of the
relations (\ref{eq3.1}) or the algebra given by Eq. (\ref{eq3.2})
in our subsequent discussions; what is required for our study is the conserved
Noether charges $H$, $D$ and $K$. We just mention, in passing, that the results
described in this note are valid for any non-relativistic theory with a
dynamical $O(2,1)$ symmetry. 

The generators $H$, $D$ and $K$ are constant in time and lead to the following
equations,
\be
\frac{d H}{d \tau} = 0, \ \
\frac{d I_1}{d \tau} = 2 I_2, \ \
\frac{d I_2}{d \tau} = H.
\label{eq4}
\ee
\noindent For time independent solutions, both $I_1$ and $I_2$ do not depend
on $\tau$. As a consequence, the energy of the static solutions of $H$ 
vanishes. This is also the case for the single-component BEC in
$2+1$ dimensions. Even though there are extra terms due to inter-component
interaction in
the case of multi-component BEC, the vanishing of the energy is a universal
consequence of the underlying $O(2,1)$ symmetry. The second equation of
(\ref{eq4}) shows that the moment $I_2$ is proportional to the time-variation
of the moment $I_1$. Recalling that the moment $I_1$ is identified as the total
mean-square radius of the condensate, the moment $I_2$ can be related to 
the speed of the growth of the condensate. This interpretation is also evident
in the definition of $I_2$ in Eq. (\ref{eq2}) after decomposing the current
${\bf j}$ as a product of the density $\rho$ and the velocity.

Defining $X=\sqrt{I_1}$, it is
easy to find a decoupled equation for $X$ from (\ref{eq4}),
\be
\frac{d^2 X}{d \tau^2} = \frac{Q}{X^3}, \ \
Q = I_1 H - I_2^2, \ \ \frac{d Q}{d \tau}=0.
\label{eq4.1}
\ee
\noindent The constant of motion $Q$ is the Casimir operator of the
$O(2,1)$ symmetry. Note that the information on the Hamiltonian $H$ is
solely contained in $Q$. Thus, the effect of the interaction, say for
example the strongly repulsive or attractive inter-component and
intra-component interaction, will be manifested through initial conditions
on $H$.  Eq. (\ref{eq4.1}) can be interpreted as the equation of
motion of a particle moving in an inverse-square potential. Interestingly
enough, this system also has a dynamical $O(2,1)$ symmetry. This reduced
system of a particle in an inverse-square potential is a
well-studied problem and the solution is given by\cite{trans},
\be
X^2 = ( a + b \tau )^2 + \frac{Q}{a^2} {\tau}^2,
\label{eq4.2}
\ee
\noindent where $a$ and $b$ are integration constants. Although any exact
solution of the equation of motion of the action ${\cal{A}}$ is not known,
it is surprising to note how the exact time-dependence of the moment $I_1$
can be obtained easily using the underlying symmetry. We would like to stress
that we are able to determine the dynamics of the total mean-square radius
of the condensate only. The dynamics of the individual mean-square radii
associated with each components can not be obtained using our method even
when there is an additional $U(1)^n$ symmetry in the system or there is no
inter-component interaction. 

The criteria for the collapse of the condensate at a
finite and real time $\tau^*$ is $Q \leq 0$. In particular, the moment $X^2$
vanishes at a finite time $\tau^*$,
\be
\tau^* = \frac{a^2}{(a^2 b^2 + Q)} \left [ - ab \pm \sqrt{-Q} \right ],
\label{eq4.2.1}
\ee
\noindent which is real if $ Q \leq 0$. Note that we have the freedom of
making $\tau^*$ either positive or negative by choosing appropriate values
for the integration constants $a$ and $b$. With the interpretation of
(\ref{eq4.1}) as a particle moving in an inverse-square potential, the
collapse of the condensate can be understood as the fall of the particle
to the center for attractive interaction. Recall that the moment $I_1$
is semi-positive definite by definition. Thus, the exact expression for
$Q$ implies that the condensate collapses for any initial condition if
$ H \leq 0$. On the other hand, if $H > 0$, the condition for the collapse
is given by,
\be
\frac{d I_1}{d \tau} \Bigm|_{\tau=0} \leq - 2 \sqrt{I_1\mid_{\tau=0} {\mid
H \mid}}.
\label{eq4.2.2}
\ee
\noindent We have used the second equation of (\ref{eq4}) in the exact
expression for $Q$ in deriving the above equation. As far as we are aware of,
this is the first instance in the literature where the criteria for the
collapse of the condensate of the most general two-dimensional multi-component
NLSE with cubic nonlinearity is given.
The criteria is independent of the total number of components $n$ and
any additional global internal symmetry. Thus, the well-known results on the
single-component\cite{za} and the multi-component\cite{tsu} NLSE
in two dimensions are easily reproduced from our general result.

Consider the following time-dependent transformation,
\bea
&& \tau \rightarrow t = t(\tau), \ \ \dot{\tau}(t) =
\frac{d \tau(t)}{d t},\nonumber \\
&& {\bf r} \rightarrow {\bf r_h} = {\dot{\tau}}(t)^{-\frac{1}{2}} \ \ 
\left ( \matrix{ {cos f(t)} & {sin f(t)} \cr
{-sin f(t)} & {cos f(t)}}
\right ) \ {\bf r},\nonumber \\
&& \psi_a(\tau, {\bf r}) \rightarrow \psi_a^h(t, {\bf r}_h) =
\dot{\tau}^{\frac{d}{4}} exp \left ( - i m \frac{\ddot{\tau}}{4 \dot{\tau}}
r_h^2 \right ) \psi_a(\tau, {\bf r}),
\label{eq5.1}
\eea
\noindent with arbitrary $\tau(t)$ and $f(t)$. Note that this transformation
can be obtained by first using the transformation (\ref{eq5}) and then a
time-dependent rotation around the z-axis with a time-dependent angle
$f(t)$. For arbitrary $\tau(t)$ and $f(t)$, the transformation (\ref{eq5.1})
is not a symmetry transformation of the action ${\cal{A}}$, instead, it maps
${\cal{A}}$ to a new action ${\cal{A}}_h= \int dt d^2 {\bf r} {\cal{L}}_h$.
The new Lagrangian ${\cal{L}}_h$ reads as,
\bea
{\cal{L}}_h & = & \sum_a \left ( i {\psi_a^h}^* \partial_{t} \psi_a^h
- \frac{1}{2 m} {\mid {\bf \grad}_h \psi_a^h \mid}^2 \right )\nonumber \\
&& - \frac{1}{2} \sum_{abcd} g_{abcd} \ {\psi_a^h}^* {\psi_b^h}^*
\psi_c^h \psi_d^h\nonumber \\
&& - \sum_a \left ( \frac{1}{2} m \omega(t) r_h^2 {\mid \psi_a^h \mid}^2
+ \dot{f} {\psi_a^h}^* L_z \psi_a^h \right ),
\label{eq5.2}
\eea
\noindent where the z-component of the angular momentum,
$L_z= - i {\bf r}_h \times \grad_h$ and the time-dependent
frequency $\omega(t)$ of the harmonic trap is determined as,
\be
\ddot{b} + \omega(t) b = 0, \ \ b(t) = \dot{\tau}^{-\frac{1}{2}}.
\label{eq5.3}
\ee
\noindent The Lagrangian ${\cal{L}}_h$ is that of a rotating multi-component
BEC in an arbitrary time-dependent harmonic well. Note that the external
harmonic potentials are identical for all the components of the condensate.
This is not by a choice. In fact, we do not have the freedom of generating
different harmonic potentials for different components using the transformation
in Eq. (\ref{eq5.1}). This is even true for the free theory, i.e. $g_{abcd}=0$.
The reason is that the transformation of the scalar fields is coupled with that
of the space-time coordinates. Consequently, unphysical and unwanted terms
will be generated in the new Lagrangian ${\cal{L}}_h$ unless all the components
of the condensate transform identically.

The solutions of ${\cal{A}}$ and ${\cal{A}}_h$ are related to
each other through the transformations in Eq. (\ref{eq5.1}) with $\tau(t)$
determined for a specific trap-frequency by the Eq. (\ref{eq5.3}). The
scale-factor $\tau(t)$ can obviously be exactly determined for a large class
of $\omega(t)$. However, the exact solutions are not known for either
${\cal{A}}$ or ${\cal{A}}_h$. This is a major problem in making use of
the mapping relating ${\cal{A}}$ to ${\cal{A}}_h$ and the vice versa.
However, note that the dynamics of the moment $I_1$ is uniquely determined
by Eq. (\ref{eq4.2}) independent of whether any exact solution of ${\cal{A}}$
is known or not. Thus, the transformation (\ref{eq5.1}) can be used to find
the dynamics of the moment
$I_{1,h} = \sum_a \int d^2 {\bf r}_h r_h^2 {\mid \psi_a^h \mid}^2$ from $I_1$.
In particular, they are related to each other by the relation,
\be
X_h = \sqrt{I_{1,h}} = b(t) X(\tau(t)),
\label{eq6}
\ee
\noindent where $b(t)$ and $\tau(t)$ are determined from the Eq. (\ref{eq5.3}).
Thus, even though the exact solution of the equation of motion of ${\cal{A}}_h$
is not known, the dynamics of $X_h$ can be described exactly.

An alternative, but, useful expression for the $X_h$ can be determined from
the following equation\cite{ami},
\be
\frac{d^2 X_h}{d t^2} + \omega(t) X_h = \frac{Q_h}{X_h^3},\
Q_h = I_{1,h} H_h - I_{2,h}^2,
\label{exact}
\ee
\noindent where $Q_h$ is a constant of motion. Both
$H_h$ and $I_{2,h}$ have the same expressions as in $H$ and $I_2$, respectively,
with $( \tau, {\bf r}, \psi_a)$ replaced by $(t, {\bf r}_h, \psi_a^h)$.
Note that the Eq. (\ref{exact}) can also be interpreted as describing the
motion of a classical particle in a combined harmonic and inverse-square
potential. The particle falls to the center for an attractive $( Q_h < 0$ )
inverse-square potential, independent of the time-dependence of the harmonic
trap. This implies that the condensate collapses at a finite time for
$ Q_h < 0$. Analyzing the exact expression for $Q_h$ further, we find that
the condensate collapses for any initial condition if $H_h \leq 0$. For
$H_h > 0$, the condition for the collapse is given by,
\be
\frac{d I_{1,h}}{d t} \Bigm|_{t=0} < - 2 \sqrt{ ( I_{1,h} \ H_h )\mid_{t=0}},
\label{criteria}
\ee
\noindent where the relation\cite{ami}, $\dot{I}_{1,h} = 2 I_{2,h}$, valid
for the system described by the action ${\cal{A}}_h$ has been used. As far
as we are aware of, this is for the first time in the literature that a
criteria for the collapse of the condensate in the most general two
dimensional multi-component GPE with cubic nonlinearity and an arbitrary
time-dependent harmonic trap is
given. Note that the criteria is independent of the total number of components
$n$ and any additional internal global symmetry. The known results
for the single-component\cite{mw} and the multi-component\cite{tsu} GPE with
time-independent harmonic trap in $2+1$ dimensions are easily reproduced from
this very general result. Further, the criteria for the collapse in the system
without or with the harmonic trap is also identical, except that there is
no equality sign in (\ref{criteria}) for the later case [ compare with
Eq. (\ref{eq4.2.2})]. This is precisely
because, for $Q,Q_h=0$, Eqs. (\ref{eq4.1}) and (\ref{exact}) describe the
dynamics of a free particle and that of a particle in a time-dependent
harmonic trap, respectively. Thus, nothing can be said conclusively on the
dynamical (in)stability for the later case, unless the time-dependence of
the frequency of the trap is explicitly specified. For a time-independent trap,
the equality sign is recovered in Eq. (\ref{criteria}); and of course, the
known result \cite{tsu,mw} is identically reproduced.

We have shown that the criteria for the collapse of the condensate
in a $2+1$ dimensional system governed by the Lagrangian ${\cal{L}}_h$
is independent of the total number of components $n$. It is known\cite{mw}
that the same criteria for the collapse of the condensate is also valid
for the Lagrangian ${\cal{L}}_h$ in dimensions $ d \geq 2 +1 $ with
$n=1$, $\omega(t)=\omega_0=$ a constant and $\dot{f}=0$. So, the criteria
for the collapse is independent of the underlying $O(2,1)$ symmetry, which
the cubic NLSE has only in $d=2+1$. The dynamical $O(2,1)$ symmetry only
helps us in deriving the exact result in a much more simpler and elegant
way. Based on this observation, we conjecture that the criteria for the
collapse of the condensate of ${\cal{L}}_h$ in dimensions $d \geq 2+1$,
with $\omega(t)=\omega_0$ and $\dot{f}=0$, is independent of the total
number of components $n$ and the criteria is the same as stated in this note
for $d=2+1$. Note that the physically interesting case of $d=3+1$ is also
included in our conjecture.

The solution for $X_h$ is given by,
\be
X_h^2 = u^2(t) + \frac{Q_h}{W^2} v^2(t), \ W(t) = u \dot{v} - v \dot{u},
\label{eq7}
\ee
\noindent where $u(t)$ and $v(t)$ are two independent solutions of Eq.
(\ref{eq5.3}) satisfying $u(t_0)=X_h(t_0)$, $\dot{u}(t_0)=\dot{X}_h(t_0)$,
$\dot{v}(t_0)=0$, and $v(t_0) \neq 0$. The above solution is valid for
arbitrary $Q_h$ : positive, negative or zero. We will be considering 
the case $ Q_h \geq 0$ from now onwards, since we have already argued
that the condensate collapses for $Q_h < 0$. We have obtained the same
expressions (\ref{eq6}) and (\ref{eq7}) in Refs. \cite{ami,mai} for the
dynamics of the width of the wave-packet of a single-component
condensate in $2+1$ dimensions. So, the results of the Refs. \cite{ami,mai}
are equally valid for the general multi-component condensate in $2+1$ dimensions
with the moment $I_1=X_h^2$ identified as the {\it total mean-square radius}.
In particular,\\
(a) the system described by ${\cal{L}}_h$ has an explosion-implosion duality
for $\dot{f}(t)=0$ and either $\omega(t)=0$ or $\omega(t)=t^{-2}$,\\
(b) the condensate exhibits extended parametric resonance for a periodic
$\omega(t)$ and arbitrary $f(t)$,\\
(c) the dynamic (in)stability of the system is independent of $f(t)$, i.e.,
same for both rotating as well as non-rotating BEC.\\
We refer the readers to Refs. \cite{ami,mai} for further details.

Finally, we conclude with the following comment. The results presented in this
note for the multi-component BEC are a generalization of what already has been
known for the single-component BEC in two dimensions. The results obtained in
both these cases are also identical with the identification of the moment $I_1$
as the {\it total mean square radius}. In particular, the dynamics of the
moment $I_1$ is determined from the same solvable equation as in the case of
a single-component BEC with all the information about the Hamiltonian encoded
into the constant of motion $Q_h$. Apart from its relevance to the ongoing
experiments on BEC, the importance of these results lie in its universality.
No matter how many components are there, or how they interact among themselves,
or even whether they are rotating or non-rotating, the dynamics of the
{\it total mean-square radius} is universally determined by the same equation.
This is indeed a counter-intuitive result and may be realized in the laboratory
in near future.

\acknowledgments{I would like to thank T. Deguchi and T. K. Ghosh for a
careful reading of the manuscript and comments. I would like to thank
J. Ieda, H. Morise, N. Uesugi, M. Wadati and especially T. Tsurumi for
useful discussions. This work is supported by JSPS.}


\begin{thebibliography}{99}
\bibitem{O21} We refer to the papers in Refs. \cite{trans} and \cite{rj}
for the application of $O(2,1)$ symmetry in physics. The models considered
in these papers also have a dynamical $O(2,1)$ symmetry.
\end{thebibliography}
\end{document}